\UseRawInputEncoding
\documentclass[%
 reprint,
 amsmath,amssymb,
 aps,prd,unsortedaddress,superscriptaddress,
 floatfix,caption,subcaption,
]{revtex4-2}
\usepackage{amsmath,amstext,amssymb}
\usepackage{float}
\usepackage[T1]{fontenc}
\usepackage{xcolor}
\usepackage{soul}
\usepackage{comment}
\usepackage{graphicx}
\setstcolor{red}
\usepackage{lineno}
\usepackage{appendix}
\usepackage[artemisia]{textgreek}
\usepackage[colorlinks=true]{hyperref}
\usepackage[doipre={DOI:~}]{uri}
\usepackage[utf8]{inputenc}
\usepackage{url}
\usepackage{natbib}

\newcommand{\appropto}{\mathrel{\vcenter{
  \offinterlineskip\halign{\hfil$##$\cr
    \propto\cr\noalign{\kern2pt}\sim\cr\noalign{\kern-2pt}}}}}

\begin{document}

\preprint{APS/123-QED}

\title{Search for gravitational-wave bursts in LIGO data at the Schenberg antenna sensitivity range}

\author{Julio C{\'e}sar Martins}
    \email[Corresponding author: ]{julio.martins@inpe.br}

\affiliation{Instituto Nacional de Pesquisas Espaciais, 12227-010 S{\~a}o Jos{\'e} dos Campos, S{\~a}o Paulo, Brazil}

\author{Ik Siong Heng}%
\affiliation{%
 SUPA, University of Glasgow, Glasgow G12 8QQ, United Kingdom
}%

\author{Iara Tosta e Melo}
\affiliation{INFN, Laboratori Nazionali del Sud, I-95125 Catania, Italy
}%

\author{Odylio Denys Aguiar}
\affiliation{%
Instituto Nacional de Pesquisas Espaciais, 12227-010 S{\~a}o Jos{\'e} dos Campos, S{\~a}o Paulo, Brazil}



\begin{abstract}

The Brazilian Mario Schenberg gravitational-wave detector was initially designed in the early 2000s and remained operational until 2016 when it was disassembled. To assess the feasibility of reassembling the Schenberg antenna, its capability to detect gravitational waves (GW) within its designed sensitivity parameters needs to be evaluated. Detection of significant signals would serve as a catalyst for rebuilding the detector. Although the antenna is currently disassembled, insights can be gleaned from the third observing run (O3) data of the LIGO detectors, given the similarities between Schenberg's ultimate sensitivity and the interferometers' sensitivity in the [3150-3260] Hz band. The search focused on signals lasting from milliseconds to seconds, with no assumptions about their morphology, polarization, and arrival sky direction. Data analysis was performed using the coherent WaveBurst (cWB) pipeline in the frequency range between 512 Hz and 4096 Hz, specifically targeting signals with bandwidths overlapping the Schenberg frequency band. However, the O3 data did not yield statistically significant evidence of GW bursts. This null result allowed for the characterization of the search efficiency in identifying simulated signal morphologies and setting upper limits on the GW burst event rate as a function of strain amplitude. The current search, and by extension the advanced version of the (a)Schenberg antenna, can detect sources emitting isotropically $5 \times 10^{-6}$ $M_{\odot}c^2$ in GWs from a distance of 10 kiloparsecs with a 50\% detection efficiency at a false alarm rate of 1 per 100 years. Moreover, we revisited estimations of detecting f-modes of neutron stars excited by glitches, setting the upper limit of the f-mode energy for the population of Galactic pulsars to $\sim 8 \times 10^{-8} M_{\odot}c^2$ at 3205 Hz. Our simulations and the defined detection criteria suggest f-modes are a very unlikely source of gravitational waves for the aSchenberg. Nevertheless, its potential in probing other types of gravitational wave short transients, such as those arising from supernova explosions, giant flares from magnetars, post-merger phase of binary neutron stars, or the inspiral of binaries of primordial black holes with sub-solar masses, remains promising.

\end{abstract}

\maketitle


\section{\label{sec:level1}INTRODUCTION
}
In recent years, Brazil has emerged as an active participant in the global landscape of experimental gravitational wave research, spearheaded by the Mario Schenberg antenna. Unlike its predecessors, which employed bar-shaped resonant-mass detectors, the Schenberg antenna features a spherical mass configuration. The main component of the detector, weighing approximately 1150 kg and measuring 65 cm in diameter, is composed of a copper-aluminum alloy, consisting of 94\% copper and 6\% aluminum \cite{Schemberg_description}. It exhibits sensitivity to signals within the frequency range of 3150 Hz to 3260 Hz. Housed within a cryogenic chamber and suspended by a sophisticated suspension mechanism, the sphere incorporates nine transducers that convert mechanical vibrations into electrical signals for subsequent data analysis and gravitational wave detection. After completing its final observational run in 2015 at a temperature of 5.0 Kelvin, the entire detector was disassembled at the University of São Paulo's Physics Institute in 2016 \cite{Schenberg_uptade}.

The Schenberg antenna requires improvements in various aspects to reach its design sensitivity.  Achieving high vacuum conditions while interfacing with cryogenics is crucial for further enhancing sensitivity \cite{dissertacao_arthur}. Upgrades in components such as lower noise amplifiers and mixers, AD (analog/digital) converters with smaller minimum voltage recording capabilities, and improved digital filters are part of the strategy. Hardware improvements related to transducers' electrical and mechanical quality factors, lower temperature operation, and better vibration isolation are also necessary \cite{Schemberg_sensitivity}. 

The prospect of detecting gravitational waves using resonant masses, which operate on a distinct physical principle compared to interferometry, holds great scientific interest. The Brazilian antenna project facilitates the expansion of gravitational wave astronomy in South America, nurturing expertise in experimental techniques and data analysis while actively engaging in international collaboration.

\begin{figure}[ht]
    \includegraphics[width=\linewidth]{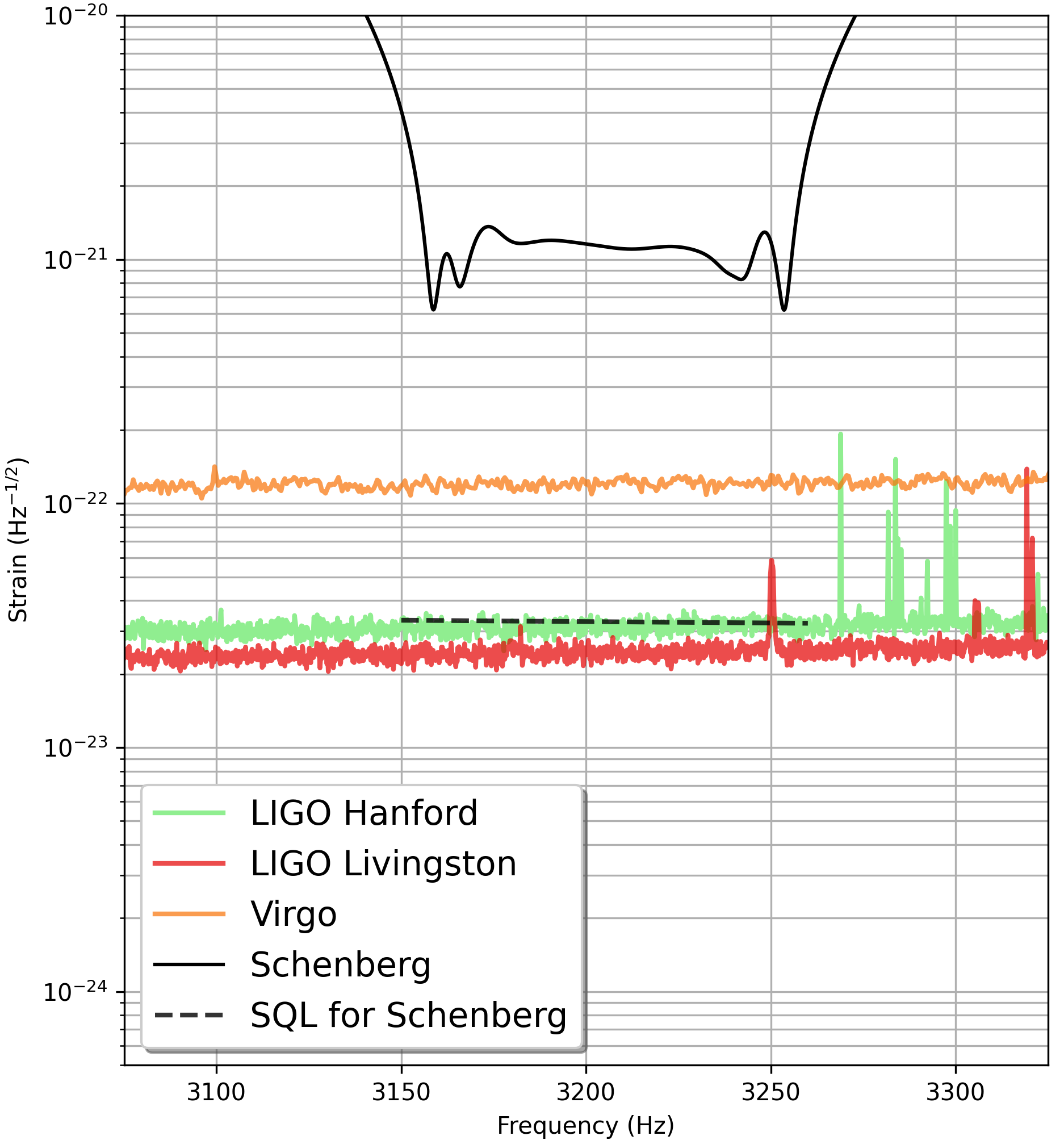} 
     \caption{The Schenberg design strain sensitivity for the overall system (not only a single mode channel), with a non-degenerate sphere, in a cryogenic system at 0.1 K \cite{Schenberg_curve} compared to the representative spectral strain sensitivity of the LIGO Hanford (29 April 2019 11:47 UTC), LIGO Livingston (5 September 2019 20:53 UTC) and Virgo (10 April 2019 00:34 UTC) during O3. }\label{fig:sensitivities} 
\end{figure}

Although the initial version of the Schenberg antenna's sensitivity curve does not match the robustness of Advanced LIGO's (aLIGO) sensitivity during the third observing run (O3) \cite{aLIGO}, an advanced version (aSchenberg) incorporating state-of-the-art electronics, exceptional mechanical and electrical quality factors, and the reduction of electronic and thermal noise can substantially narrow the gap. The aSchenberg configuration aims to approach the standard quantum limit, a critical stage in achieving ultimate sensitivity. Notably, the sensitivity goals of the aSchenberg detector have already been attained by the two LIGO interferometers during the O3 run, as depicted in Figure \ref{fig:sensitivities}. Consequently, analyzing the O3 LIGO data within the Schenberg antenna's bandwidth offers a valuable opportunity to evaluate the potential for gravitational wave detection with the aSchenberg detector operating at its ultimate sensitivity.

The detection of GW signals in LIGO data at Schenberg sensitivity could catalyze the reassembly and upgrade of the Brazilian antenna. At its ultimate sensitivity, the Schenberg antenna could join the ground-based interferometers in a GW detector network. 
Unlike interferometers, spherical antennas, characterized by a particular arrangement of transducers \cite{TIGA}, have an omnidirectional response, making them equally sensitive to gravitational waves of all polarizations and directions \cite{omnidirectional_sphere}. A single spherical detector can provide source direction estimates with reasonable resolution \cite{Inverse_problem_sphere}. Additionally, the antennas can measure all tensorial components of a GW, providing valuable information to test alternative theories of gravitation \cite{Test_GR_sphere}.
This synergy between ground-based interferometers, which are better at reconstructing GW waveforms due to their broader bandwidth, and spherical antennas, which are better at reconstructing GW arrival directions \cite{Inverse_problem_sphere}, could enhance our understanding of GW sources. Further, this combination results in a network with interesting features for multi-messenger astrophysics \cite{Multimessenger}.

The Compact Binary Coalescence (CBC) signals detected by ground-based interferometers so far \cite{GWTC-1, GWTC-2, GWTC-2.1, GWTC-3} were not observable in this narrow frequency band, given the masses involved. The last stable orbit (LSO) of these binary systems prevents the frequency from increasing indefinitely when stars begin to interact and merge, corresponding to a typical value of $f_{LSO} \sim 220 (20M_{\odot}/M)$ Hz for a compact binary with total mass $M$ \cite{GW_review}. Although the characteristics of the inspiral and merger processes limit high-frequency components of the signal other cosmic events may produce signals within the Schenberg band.  The spherical antenna could capture burst signals from fundamental oscillation modes (f-modes) of neutron stars, the excitation of the first quadrupole normal mode of 4-9 solar mass black holes \cite{Schenberg_events}, and the post-merger phase of binary neutron star coalescence \cite{postmerger_NS}.

This investigation aims to assess the detectability of gravitational wave signals using the Schenberg spherical antenna at its \textit{ultimate} sensitivity. The analysis is intentionally designed to explore a wider frequency range that encompasses the Schenberg bandwidth ($\Delta f_{Sch}$), enabling the search for potential gravitational wave emissions close to this band. Utilizing the coherent WaveBurst algorithm \cite{cwb_soft}, we conducted a comprehensive search for gravitational wave burst signals within the O3 data collected between April 1st, 2019, and March 27th, 2020. 
Unlike searches focusing on specific sources, our all-sky approach allows for the exploration of diverse signal morphologies and enhances the search sensitivity to a broad variety of sources \cite{A_Guide_Transient}.

In line with our approach, we adhere to a well-established methodology employed in the search for gravitational-wave bursts~\cite{Injections, Allsky_burst_O1, Allsky_burst_O2, Drago_thes, sigmoid_burst, first_joint_run}. Our focus centers on the most recent investigation conducted on the LIGO and Virgo data~\cite{O3_allsky_burst}. In Section~\ref{sec:DATA SET}, we provide a concise overview of crucial aspects regarding the LIGO data utilized in the search, including access protocols, data analysis limitations, and its relevance to our investigation. Section~\ref{sec:GWs burst search} comprehensively outlines the features of the search pipeline, highlighting its primary detection parameters. The data analysis and main findings are presented into background and foreground events, facilitating the estimation of detection statistics. Subsequently, in Section~\ref{sec:Astrophysical_interpretation}, we delve into astrophysical interpretations of the results, shedding light on various aspects such as the efficiency of the cWB method, upper limits on burst rates, and the detectability estimation of f-modes in neutron stars. Finally, in Section~\ref{sec:conclusion}, we conclude our paper by outlining the implications of our findings in relation to the Schenberg antenna, encapsulating the significance of our research within the broader context of gravitational wave studies.

\section{Data set}
\label{sec:DATA SET}

The data from the Third Observational Run (O3) were collected from April 1, 2019, to March 27, 2020, and included the two LIGO detectors~\cite{LIGO_O3} and Virgo's participation~\cite{Virgo_O3}. From October 1, 2019, to November 1, 2019, a commissioning break split O3 into two large parts, the first one containing six months of data (O3a) and the second part known as O3b, which all together corresponds to 330 days of observational run. The analysis presented here is based on both parts of the O3 data set.

The coherent analysis of the data requires the participation of more than one detector. Especially for high frequencies, Virgo had a considerably higher noise floor than the LIGO detectors for O3. Therefore, Virgo's participation in the coherent analysis does not improve the selection of coincident events, while the high rate of non-Gaussian noise would increase the overall false candidates. In order to maximize the chance of detecting real GW events, we use only the Hanford-Livingston (HL) network, in agreement with other works on coherent search for unmodeled signals in O3~\cite{O3_allsky_burst, O3_allsky_long_burst}. 

During the O3a run, the duty cycle, i.e., the amount of time in the run that the instruments were effectively observing, was $71 \%$ (130.3 days) for LIGO Hanford and $76 \%$ (138.5 days) for LIGO Livingston. For O3b, the duty cycles were $79 \%$ (115.7 days) for LIGO Hanford and $79 \%$ (115.5 days) for LIGO Livingston~\cite{LIGO_characterization}. All the GW strain O3 data used are available at the Gravitational Wave Open Science Center (GWOSC) \cite{gwosc_paper} sourced from the channel \textit{DCS-CALIB\_STRAIN\_CLEAN-SUB60HZ\_C01} \cite{GWOSC_O3}. 

The amount of analyzed data is reduced due to the requirement of coincident observation of the HL network and the removal of poor-quality segments of each detector's data stream, a total of 102.5 days of data for O3a and 93.4 days for O3b. The list of data quality vetoes with the fraction of removed data and their respective noise sources for O3 that was considered in this work is available at~\cite{O3dataflags}.

The Advanced LIGO data is calibrated to detect signals with frequencies from 10 Hz to 5 kHz and out of this bandwidth does not have either uncertainty characterization or assigned reliability. Some works characterized the calibration uncertainty of the data during O3a~\cite{uncertainty_O3a} and O3b~\cite{uncertainty_O3b} and the full files of LIGO calibration uncertain are available to public access at~\cite{O3calibration_file}. The systematic error bounds are not expected to impact this search for gravitational-wave bursts. In the frequency range of 512-4096 Hz, the upper limits on the systematic errors and the associated uncertainty of LIGO detectors data in O3 are $<$17$\%$ in magnitude and $<$12º in phase. Also, the estimation of the network timing uncertainty ($\sim$10 $\mu s$) could be neglected when compared to the uncertainty in estimates of the time-of-arrival for a GW event ($\sim$1 $ms$)~\cite{uncertainty_O3a}.

\section{gravitational-wave bursts search}
\label{sec:GWs burst search}
\subsection{\label{subsec:cwb}coherent WaveBurst pipeline}

The coherent WaveBurst (cWB) is an open-source software for gravitational-wave data analysis~\cite{cwb_soft}. From the strain data of the detector network, the cWB search pipeline uses a coherent analysis \cite{coherent_analysis} to identify and reconstruct transient signals of gravitational waves without prior knowledge of signal morphology. With the computational core developed in the $C{++}$ programming language, the cWB is one of the main pipelines used by the LIGO-Virgo-Kagra (LVK) Scientific Collaboration, working on low-latency during the observing runs and in the offline searches~\cite{cwb}. It is the only algorithm sensitive to generic morphologies in the all-sky search used in gravitational-wave transient catalogs.  

The cWB uses a method that combines the individual data streams of a multi-detector network in a coherent statistic based on the constrained maximum likelihood analysis that allows the reconstruction of the source sky location and the signal waveform. In this framework, despite searching for non-specific morphologies, the cWB can effectively discern authentic GW signals from those caused by transient noises. A more detailed description can be found at~\cite{cwb_analit}. 

The data are represented in the time-frequency domain by Wilson-Daubachies-Meyer (WDM) wavelet transform~\cite{wavelet}, expressing the data through "pixels" with different bases and resolutions. From this parameterization, the data are whitened, and those pixels that exceed a pre-established energy value are selected and clustered with similar ones from their neighborhood. 
Under the assumption of Gaussian white noise, the clusters of selected pixels are evaluated by the logarithmic value of the likelihood ratio with a functional form that takes account of the source sky directions, the antenna pattern functions, and the time delay of the signal between the detectors in the network~\cite{cwb_analit}. The cWB's detection statistics are based on the maximum likelihood ratio framework, split into incoherent and coherent parts, and are used to evaluate a significant cluster of pixels designated \textit{trigger}.

In the cWB algorithm, the null energy or residual noise energy ($E_n$) is calculated as the sum of the squared amplitudes of the elements in the null stream, first introduced by G\"{u}rsel and Tinto \cite{null_stream}, which is constructed by combining data from multiple detectors to cancel out coherent gravitational wave signals while retaining information about noise artifacts. The null energy employs the different sensitivities and noise characteristics of the detectors in the network to provide a measure of the inconsistencies between the reconstructed signal and the cluster selected from detector outputs, allowing for effective discrimination between true gravitational wave signals and transient noise or \textit{glitches} \cite{glitches_S6,glitches_O1,LIGO_characterization}. 
 
Contrastingly, the coherent energy ($E_c$) statistics depend on the cross-correlation terms of the reconstructed waveform in different detectors.
Together, these energy statistics are used to define two key parameters of the trigger.
One is the network correlation coefficient $c_c$, an efficient parameter to distinguish genuine GW signals from those caused by accidental coincident noises in the detectors. It is quantified by:

\begin{equation}
    c_c \equiv  \frac{E_c}{|E_c| + E_n}, 
 \label{correlation_coef}
\end{equation}

\noindent where gravitational-wave authentic signals are expected to have $c_c \approx 1$ and coincident glitches $c_c \ll 1$.
The main burst detection statistic used by cWB is:

\begin{equation}
    \eta_c = \sqrt{\frac{c_c E_c K}{K-1}},
 \label{coherent SNR}
\end{equation}

\noindent where K is the number of detectors used in the cWB analysis. This equation gives us a parameter equivalent to the coherent network signal-to-noise ratio (SNR).

\subsection{\label{subsec:analys}Analysis procedure}

The primary analysis is based on the search setup of cWB employed in the all-sky search for short gravitational-wave bursts in O3 strain data \cite{O3_allsky_burst}. We utilize the same pre-production configuration of cWB that was used in their high-frequency analysis, covering the frequency range of 512-4096 Hz. For each trigger generated in the search, cWB establishes a parameter tree that contains various information derived from the identified power excess in the coherent data. This includes parameters such as bandwidth, central frequency, duration, network correlation coefficient, coherent network SNR, and others. 

Candidate events are selected by setting thresholds on these trigger parameters during the post-processing stage of cWB. To assess the significance of burst candidate events, their detection parameters are compared against the population of background triggers. These trigger populations are selected if their central frequency $f_0$ falls within the Schenberg band ($f_0 \in \Delta f_{Sch}$) or if their bandwidth $\Delta f$ partially overlaps with the Schenberg band ($\Delta f \cap \Delta f_{Sch} \neq \emptyset$). The bandwidth $\Delta f$ is determined from the maximum and minimum frequency information obtained from the corresponding wavelet map. In our analysis, we adopt a network correlation coefficient threshold of $c_c = 0.8$, bounding $\eta_c$. 

While the analysis does not precisely emulate the sensitivity range of the aSchenberg detector, narrowing the search to only signals within the Schenberg band could lead to an inaccurate assessment of the detector's real detection potential.  By considering signals in a wider frequency range, we gain valuable insights into the occurrence frequency and distribution of burst events with origins potentially linked to signals within the Schenberg frequency range. Signals detected by the interferometers, centered outside the [3150-3260] Hz range, may still appear as triggers for Schenberg, but with lower energy levels. Combining these data could particularly improve the high-frequency components of the strain data allowing for a more detailed investigation of their characteristics and the properties of astrophysical sources. 

The assessment of the significance of candidate events relies on the specific set of thresholds applied to the parameters of the trigger population. Although analyzing the triggers generated by another search with the same configuration, it is important to note that we do not anticipate obtaining the same statistical significance results.    

\subsubsection{Background analysis}

To accurately assess the significance of burst candidate events, it is crucial to understand the statistical properties of the coincident non-Gaussian noise present in the detector network. These noise sources arise from random instrumental artifacts within each interferometer, resulting in coincident transients that can resemble gravitational-wave signals. When identified appropriately, the coincident noise events are referred to as background triggers.

The generation of background samples involves a time-shift procedure. Specifically, the data stream from one detector is shifted in time relative to the other by a time interval exceeding the coincidence window, $\sim$10 ms \cite{GW_network}. By applying this time-lag to the data streams, triggers that are identified through network coherent analysis are considered fortuitous and are classified as part of the background sample. The intentional time shift eliminates the possibility of these triggers being caused by actual gravitational-wave signals, enabling us to isolate and quantify the contribution of instrumental artifacts. This process is repeated for different time lags, and the combined results provide a more robust characterization of the background triggers \cite{background} 

The non-regular distribution of background event counts required the division of the O3a analysis into two separate data chunks. Chunk 1 spans from the beginning of O3 until May 16, 2019, and has a significantly higher count of events than the rest of O3a, named chunk 2. This discrepancy persists even after applying data quality cuts, indicating a period during which noise has a more pronounced impact in the high-frequency band. Chunk 3 encompasses the entire O3b epoch.

\begin{figure}[ht]
    \includegraphics[width=\linewidth]{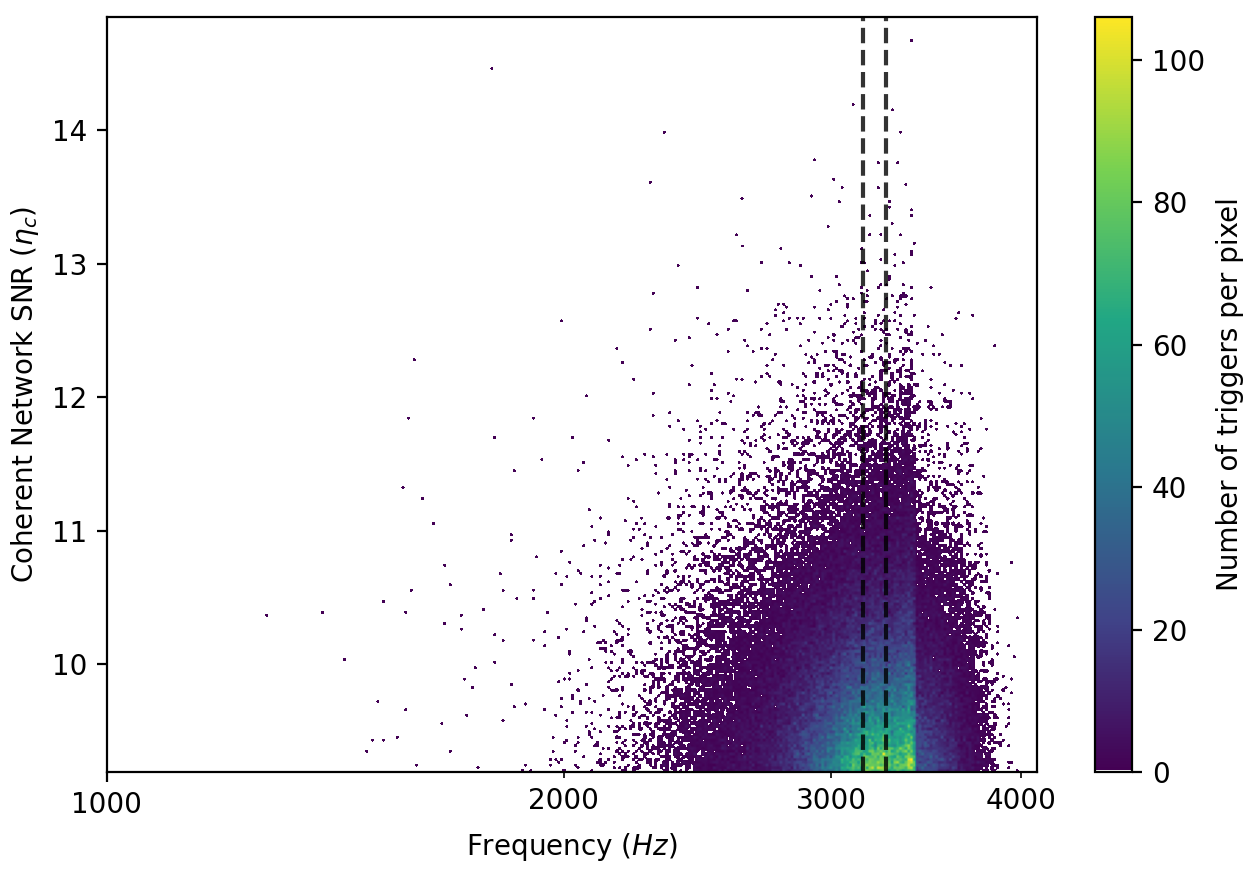}
    \caption{The distribution of coherent background triggers in O3, represented by the coherent network SNR plotted against their central frequencies. A substantial number of triggers are concentrated at the lower end of the coherent network SNR values. The variations in density along the horizontal axis are due to the additional frequency cuts applied to chunk 1, particularly the discontinuity at 3400 Hz. The vertical dashed lines indicate the frequency range corresponding to the Schenberg sensitivity band.}\label{BKG_distribution_freq} 
\end{figure}

In chunk 1, the distribution of the coherent network signal-to-noise ratio (SNR) for the trigger rate deviates from the expected Poisson behavior. The background triggers exhibited a SNR dependence on the central frequency, particularly for triggers with $f_0 > 3400$. Consequently, triggers above this central frequency were excluded from the analysis, following a similar procedure adopted by \cite{O3_allsky_burst}. Furthermore, two exceptional triggers were identified, characterized by extreme loud glitches with $\eta_c > 42$ and $f_0 < 896$ Hz. These outliers deviated significantly from the overall population of background triggers in O3, as shown in Figure \ref{BKG_distribution_freq}. To maintain the integrity of the data set, an additional selection criterion was introduced, excluding triggers with $f_0 < 896$ Hz, effectively eliminating the impact of these outliers on the analysis. 

These frequency thresholds are unique to chunk 1 due to its discrepant behavior and do not apply to the other data chunks. Consequently, the analysis of O3 is divided into three distinct and temporally separated large data blocks. Considering these characteristics, the estimated background livetime for chunk 1 is 296.2 years, while chunk 2 amounts to 258.3 years. Chunk 3 is evaluated over a period of 569.5 years.

\subsubsection{Zero-lag analysis}
\label{subsubsec:zero-lag_analysis}

The significant zero-lag coincident triggers are short-duration transients that are candidates for gravitational-wave burst events. They were selected using the same threshold on trigger parameters that was used in the background. The zero-lag livetime of the first half of O3 was 21.7 days for chunk 1, 80.8 days for chunk 2, and the second half of O3 was 93.4 days.

\begin{figure}[ht]
    \includegraphics[width=\linewidth]{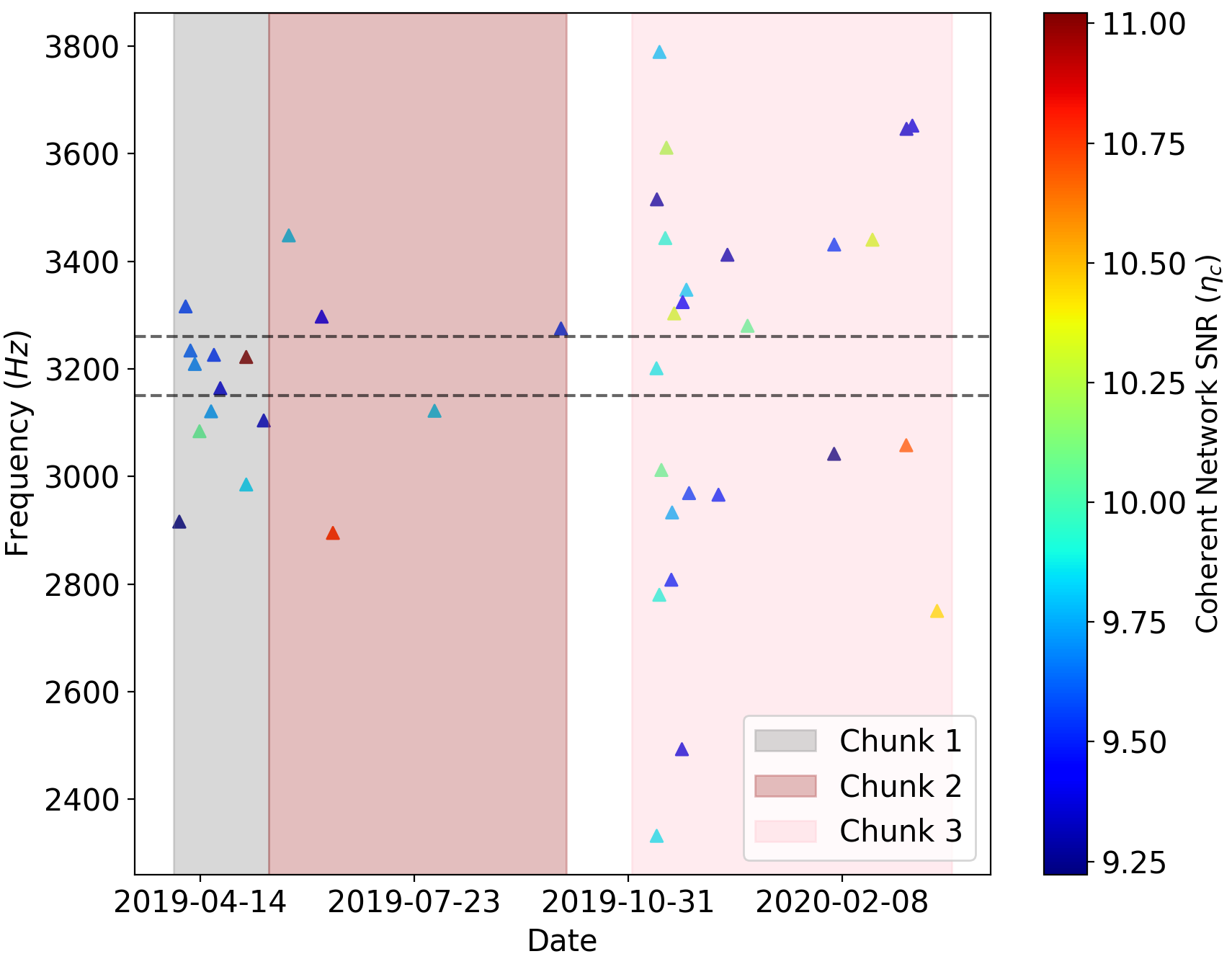}
    \caption{The plot illustrates the time versus central frequency distribution of zero-lag triggers and the corresponding coherent network SNR calculated by the cWB algorithm. The heat map visually represents the SNR values. The shaded regions in the plot demarcate the time boundaries of each chunk. The Schenberg sensitivity band is illustrated by the horizontal dashed lines.}\label{0lag}
\end{figure}

After the application of post-production thresholds, forty-one cWB triggers that were not vetoed have persisted. Out of these, eleven were found in chunk 1, five in chunk 2, and twenty-five in O3b. Figure~\ref{0lag} systematically illustrates these findings, showcasing the time, central frequency, and coherent network SNR information for each trigger. 

The loudest event has a central frequency at 3222 Hz, $\eta_c$ = 11.0, and occurred on May 05, 2019, corresponding to chunk 1 epoch.

\subsubsection{Detection statistics}
\label{subsubsec:detection_stats}

To accurately differentiate between triggers resulting from genuine gravitational wave signals and those arising from transient noise, the significance of each candidate event is assessed by considering its behavior in relation to the background noise. This assessment is accomplished through the calculation of the False Alarm Rate (FAR)~\cite{FAP}.

The FAR is a metric assigned to each zero-lag trigger, which are now referred to as foreground triggers, when a background trigger population is characterized and ranked based on an intermediate detection statistic. In the case of the coherent WaveBurst (cWB) algorithm, this ranking statistic is the coherent network signal-to-noise ratio ($\eta_c$). The significance of the foreground triggers was estimated by comparing them to the corresponding chunks of the background data. 

For every candidate event, the FAR is determined through the equation:

\begin{equation}
		 FAR = \frac{N}{\sum\limits_{i}^{}T_i} , \\ 
\label{FAR_eq}	
\end{equation} 

\noindent where $N$ represents the total number of background triggers with an intermediate statistic greater than or equal to the candidate event, and $T_i$ denotes the analyzed duration in the $i$-th chunk \cite{FAR_thesis}. When the FAR value is low, it means that there is less chance that the event was caused by detector noise. To make this information more understandable, the iFAR (inverse False Alarm Rate) is often used by simply taking the inverse of the FAR. The estimation of background, with different thresholds and trigger rates, will lead to varying levels of statistical significance for the foreground triggers.

\begin{figure}[ht]
    \includegraphics[width=\linewidth]{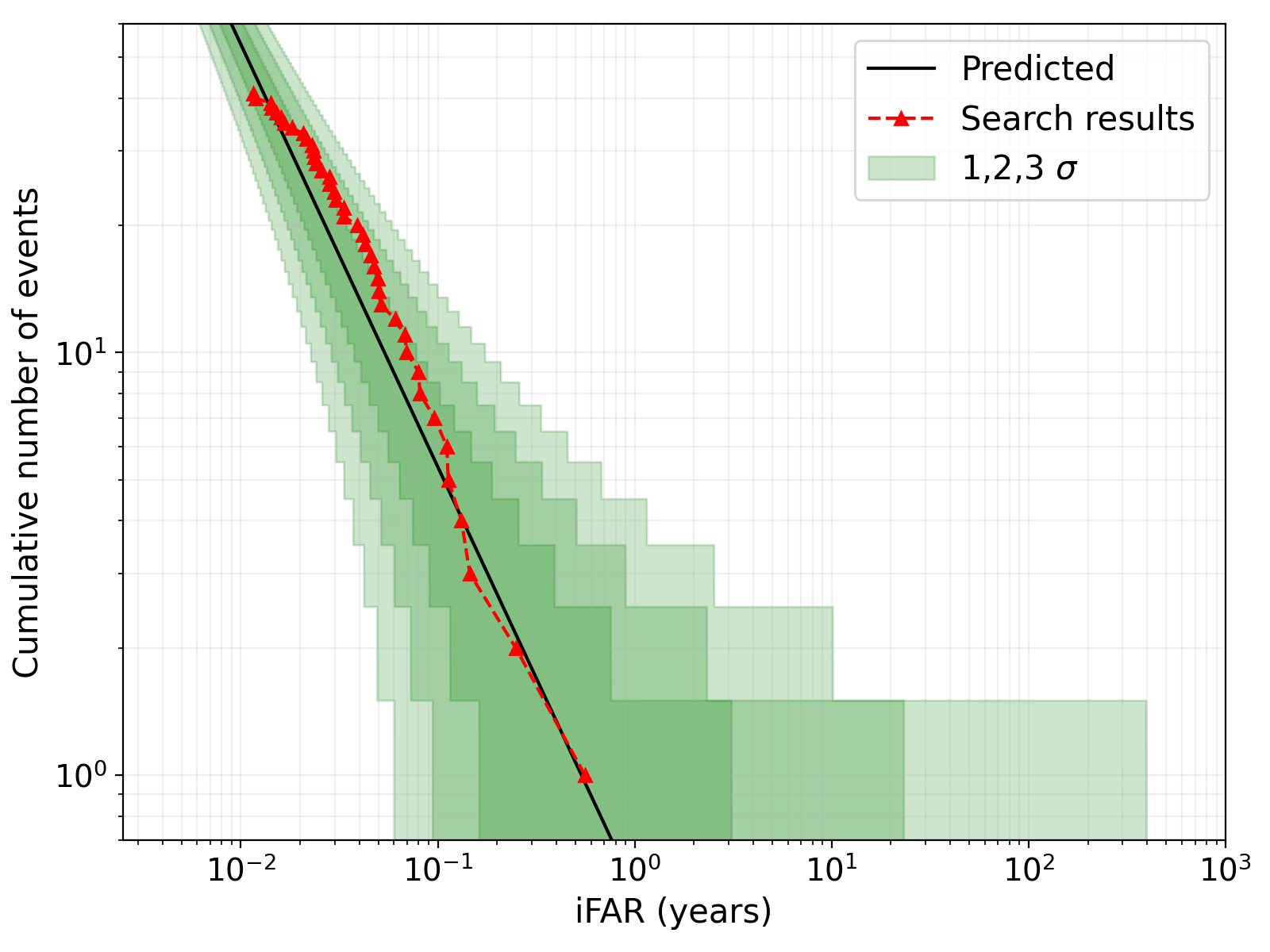}
    \caption {The cumulative number of zero-lag coincident events (foreground), with some measurable energy within the Schenberg band during O3, versus their significance. The green area shows the one, two, and three-standard deviation on the expected value of the background estimate significance (solid black line), assuming a Poisson distribution.}.
    \label{IFAR}
\end{figure}

Accidental coincidences, such as background triggers, follow an independent random distribution that aligns with a Poisson distribution in experimental settings. By conducting an analysis of zero-lag coincident events within the statistical window of the expected background significance, guided by the False Alarm Rate, it becomes possible to identify whether candidate events are caused by independent random noise. Figure \ref{IFAR} illustrates one approach to this type of analysis, depicting the cumulative number of triggers as a function of iFAR. For each trigger, the expected number of background events louder than the given trigger is calculated as $\Sigma_i T_i/$iFAR.

Deviations beyond the regions of Poisson probability in the iFAR values serve as indications of potential burst detection candidates. Also, we set an iFAR threshold of 100 years to identify a significant detection, consistent with other searches for gravitational-wave bursts \cite{Allsky_burst_O1, Allsky_burst_O2, O3_allsky_burst}.

This search identified as the most significant candidate event the trigger in chunk 2 with $\eta_c$ = 10.7 and iFAR = 0.6 years. This event is twice as significant as the most prominent event in the high-frequency analysis conducted in reference \cite{O3_allsky_burst}. However, none of the identified events here were significant enough to be classified as gravitational-wave bursts. All the detected events match the expected background at a three-sigma level, which is consistent with previous findings also reported in \cite{O3_allsky_burst}.

\section{Astrophysical interpretation}
\label{sec:Astrophysical_interpretation}
\subsection{Search sensitivity}
\label{subsec:sensitivity}

Astrophysical interpretations can be derived from the analysis, even in cases where gravitational-wave burst signals do not reach the statistical significance required for detection. These interpretations are based on evaluating the efficiency of the cWB pipeline in reconstructing burst signals, accomplished through a Monte Carlo method. In this procedure, synthetic gravitational-wave signals, referred to as \textit{injections}, are inserted into the LIGO data stream, and the cWB pipeline is activated to detect these injected signals, using the same configuration as the primary search. These injections are software-generated, designed to simulate the passage of gravitational-wave bursts through the LIGO detector network. The cWB incorporates the MDC Engine, an integrated burst waveform generator that offers customizable options for specifying waveforms, sky distribution, polarization, injection rate or time, and amplitude of the injections \cite{cwb_soft}.

The strength of the injections is typically quantified by the root-sum-squared strain amplitude ($h_\text{rss}$), which is determined by integrating the squared values of the two signal polarizations ($h_+$ and $h_\times$) over the entire signal duration, 

\begin{equation}
		 {h}_{rss} = \sqrt{\int^{\infty}_{-\infty} (h^2_{+}(t) + h^2_{\times}(t))dt}.
	\label{hrss}	
	\end{equation} 

This is interpreted as the amplitude of the gravitational wave that reaches Earth before being modified by the detector's antenna pattern~\cite{hrss}. The $h_\text{rss}$ is expressed in units of Hz$^{-1/2}$, allowing for a direct comparison with the spectral strain sensitivity of the detectors. Various strain factors are applied to appropriately scale the defined amplitudes for each simulated injected signal, thereby improving the assessment of cWB's efficiency.

The \textit{ad hoc} injections may not come from specific astrophysical models, but they can still approximate the morphology of gravitational-wave bursts that are expected to occur within the Schenberg frequency band. These bursts are made up of circularly polarized Sine-Gaussian (SG) and Ring-Down (RD) signals, with central frequencies that range from 3150 Hz to 3260 Hz. The waveforms are parameterized as

\begin{gather}
 \begin{bmatrix} h_{+}(t) \\ h_{\times}(t) \end{bmatrix}
 = A \times 
   \begin{bmatrix} \frac{1+\alpha^2}{2}  \\ \alpha \end{bmatrix}
    \times
  \begin{bmatrix} H_{+}(t) \\ H_{\times}(t)   \end{bmatrix},
 \label{inj_par}
\end{gather}

\noindent where the strain amplitude (A) is multiplied by the waveforms (H$_{\times/+}$(t)) representing the two independent polarizations. The parameter $\alpha$ denotes the ellipticity of the signal polarization, with a value of 1 representing circular polarization, which corresponds to an optimally oriented source~\cite{Injections}. For sources that involve rotational motion around an axis, ellipticity is defined as the cosine angle between the rotational axis of the source and the line of sight from Earth. 
The injected waveforms are distributed evenly across the sky, with nine strain factors covering a grid of $h_\text{rss}$ values ranging from $5.00\times 10^{-23}$ Hz$^{-1/2}$ to $4.05 \times 10^{-21}$ Hz$^{-1/2}$, with logarithmically spaced values stepping by $\sqrt{3}$.

Unlike the main search, which considered signals with a significant part of energy beyond the 3150-3260 Hz band, our simulation specifically focused on injections with a central frequency within the Schenberg band. This choice allows us to accurately assess the efficiency of a burst search in this frequency range, providing valuable insights into the detection capabilities of the Schenberg detector.

\subsubsection{\textit{Ad hoc} waveforms set}
\label{subsubsec:waveform_set}

The Ring-Down waveform is defined by three main attributes: the damping time denoted as $\tau$, the central frequency $f_0$, and polarizations, which are parameterized as 

\begin{equation}
\begin{split}
		 {H}_{+}(t) = \exp({-t/\tau}) sin(2\pi f_0 t), \\
		 {H}_{\times}(t) = \exp({-t/\tau}) cos(2\pi f_0 t).
\end{split}\label{RD_eq}
\end{equation}

These injections can imitate the anticipated morphology of gravitational-wave bursts that originate from quasi-normal modes in neutron stars (NSs). This approximation is particularly relevant to the fundamental modes (f-modes) associated with events like pulsar glitches or magnetar giant flares~\cite{NS_f-modes}.  Under the assumption of soft Equations of State (EoS), f-modes can exhibit frequencies of up to 3 kHz with a typical damping time of around 100 ms~\cite{f_modes_EOS}. The post-merger neutron star (PMNS), a scenario in which a massive, differentially rotating neutron star emerges, is another potential source with this signal morphology. Short gravitational wave bursts ($\sim$$10-100$ ms) with ringdown-like waveforms and dominant oscillation frequencies ($\sim$$2-4$ kHz), connected to quadrupole oscillations in the fluid, result from non-axisymmetric deformations in the NS remnant when PMNS sustain the collapse~\cite{PMNS}. Based on these characteristics, three Ring-Down waveforms were injected with damping times $\tau$ of 5 ms, 50 ms, and 100 ms, centered at 3205 Hz. Additionally, two other injections with $\tau$ = 100 ms were performed, with central frequencies located at the edges of the Schenberg band, specifically 3150 Hz and 3260 Hz.

The Sine-Gaussian waveforms are defined by their polarization, central frequency, and the quality factor parameter Q, which represents the ratio of the central frequency to the bandwidth. These Sine-Gaussian injections cover a wide range of burst specifications by varying the quality factor value. The waveforms can be expressed as shown in Equation \ref{SG_eq} 
\begin{equation}
\begin{split}
		 {H}_{+}(t) = \exp\left(-{\frac{2 t\pi^2 f_0^2}{Q^2}}\right) sin(2\pi f_0 t), \\
		 {H}_{\times}(t) = \exp\left(-{\frac{2 t\pi^2 f_0^2}{Q^2}}\right) cos(2\pi f_0 t),
\end{split}\label{SG_eq}
\end{equation}

\noindent where ${H}{+}(t)$ and ${H}{\times}(t)$ represent the two independent polarizations.  
The set of Sine-Gaussian injections comprises nine waveforms, with three different quality factor values (3, 9, and 100), each having three central frequencies that represent the start, middle, and end of the Schenberg band.

\subsubsection{Efficiency analysis}
\label{subsubsec:eff_analysis}

We evaluate the cWB pipeline's ability to detect and reconstruct injected signals to provide a reliable approach to assess the potential for burst detection by the aSchenberg antenna. This quantifies the search efficiency, which is the fraction of injected signals that can be successfully detected irrespective of their origin. The analysis considers a hypothetical population of sources represented by \textit{ad hoc} waveforms with distinct $h_\text{rss}$ values. Also, the evaluation is sensitive to various factors such as waveform characteristics, injected central frequency, pipeline configuration, and trigger parameters.

The detection efficiency is a function of the strain amplitude, as signals with higher $h_\text{rss}$ are more likely to be detected. The choice of scaling factor amplitude is established to ensure that the efficiency curve covers a range from 0\% to close to 100\% detection efficiency. Table~\ref{injections_hrss} displays the values of $h_\text{rss}$ that achieved 10\%, 50\%, and 90\% detection efficiency.

The injected signals underwent the same post-production thresholds as those applied to the foreground and background search to ensure this result reflects the performance of the search algorithm properly. Furthermore, only injections that met the criterion for significant detection (iFAR $\geq$ 100 years) were considered in the analysis.

The efficiency results presented here are differentiated only by the O3a and O3b epochs. We accounted for the significant variation in the background levels across different chunks by applying the false alarm rate (FAR) threshold to each chunk individually. This ensure that the injected signals were appropriately weighed by the corresponding background population. It also enables comparisons with other searches utilizing O3 data, as they typically differentiate between the O3a and O3b epochs.

\begin{table}[ht]
\caption{\label{injections_hrss}%
Values of $h_\text{rss}$ in units of 10$^{-22}$ Hz$^{-1/2}$ for 10\%, 50\% and 90\% detection efficiency in O3a and O3b at FAR threshold of 1/100 years.
}
\begin{ruledtabular}
\tiny
\begin{tabular}{ccccccccc}
& \multicolumn{2}{c}{$h_\text{rss}^{10\%}$} & \multicolumn{2}{c}{$h_\text{rss}^{50\%}$} & \multicolumn{2}{c}{$h_\text{rss}^{90\%}$} \\
\textbf{Morphology}   & \textbf{O3a}   & \textbf{O3b}  & \textbf{O3a}      & \textbf{O3b}     & \textbf{O3a}      & \textbf{O3b}     \\ \hline
\textbf{Ring-Down damped oscillation  (circular)} & \multicolumn{2}{c}{\textbf{}}        & \multicolumn{2}{c}{\textbf{}}        & \multicolumn{2}{c}{\textbf{}}        \\ \hline
$f_0 = 3205$ Hz, $\tau = 5$ ms   & 3.0 & 3.2 & 4.8 & 4.8 & 12.5 & 10.3\\
$f_0 = 3205$ Hz, $\tau = 50$ ms  & 2.9 & 3.1 & 4.5 & 4.7 & 9.7 & 9.3 \\
$f_0 = 3150$ Hz, $\tau = 100$ ms & 2.8 & 3.0 & 4.3 & 4.4 & 9.4 & 9.0 \\
$f_0 = 3205$ Hz, $\tau = 100$ ms & 2.9 & 3.1 & 4.4 & 4.5 & 9.6 & 9.2 \\
$f_0 = 3260$ Hz, $\tau = 100$ ms & 3.0 & 3.2 & 4.8 & 4.9 & 10.3 & 9.8 \\ 
\textbf{Sine-Gaussian wavelets (circular)} &          &          &          &          &          &          \\ 
$f_0 = 3150$ Hz, $Q = 3 $  & 3.3 & 3.5 & 5.6 & 5.4 & 20.6 & 17.3 \\
$f_0 = 3194$ Hz, $Q = 3 $   & 3.4 & 3.4 & 5.6 & 5.2 & 28.7 & 14.5 \\
$f_0 = 3260$ Hz, $Q = 3 $  & 3.6 & 3.6 & 5.9 & 5.4 & 37.9 & 15.2 \\
$f_0 = 3150$ Hz, $Q = 9 $  & 3.0 & 3.2 & 4.9 & 4.8 & 15.3 & 10.8 \\
$f_0 = 3205$ Hz, $Q = 9 $  & 3.1 & 3.2 & 5.0 & 4.9 & 14.0 & 11.2 \\
$f_0 = 3260$ Hz, $Q = 9 $  & 3.2 & 3.3 & 5.2 & 5.0 & 16.3 & 12.3 \\
$f_0 = 3150$ Hz, $Q = 100 $  & 2.7 & 2.9 & 4.3 & 4.5& 9.5 & 9.0 \\
$f_0 = 3194$ Hz, $Q = 100 $  & 2.7 & 2.9 & 4.3 & 4.5 & 8.5 & 8.9 \\
$f_0 = 3260$ Hz, $Q = 100 $  & 3.0 & 3.2 & 4.5 & 4.9 & 9.5 & 9.3 \\ 
\end{tabular}
\end{ruledtabular}
\end{table}

The comparison of $h_\text{rss}$ values between O3a and O3b, as presented in Table~\ref{injections_hrss}, reveals consistent results without significant deviations. The cWB sensitivity is influenced by waveform properties and the excess power observed in the time-frequency plane and presents better results for Sine-Gaussians with high-Q and Ring-Downs with high-$\tau$.
It performs better when the signals have a narrow frequency bandwidth, allowing for a more precise description using wavelet representations.

\subsection{Detection range}
\label{subsec:detection_range}

The connection between the amplitude of the injected waveforms and the detection efficiency allows for the study of astrophysical implications. A comprehensive characterization of the search sensitivity can be achieved by relating the results obtained from the injected amplitude grid to the GW emitted energy of potential astrophysical sources. Assuming a standard-candle source at a distance of $d_0 =$ 10 kpc this energy can be calculated using the equation \cite{Energy_isot}: 

\begin{equation}
E^{iso}_{GW} = \frac{\pi^2 c^3}{G} f_{0}^2 d_0^2 h_{rss}^2.
\label{energy_iso}
\end{equation}

By connecting $h_\text{rss}$ with a specific efficiency value obtained from the simulation study, Equation~\ref{energy_iso} provides an estimate of the minimum amount of energy that needs to be isotropically radiated by the gravitational-wave source to be detected by cWB. For rotating systems emission, expected to produce circular polarization, the energy is given as $E^{rot}_{GW} = (2/5) E^{iso}_{GW}$ \cite{Energy_isot}. It is important to note that these calculations are based on the assumption of isotropic injections distributed over the sky, which means they provide only approximate estimations. For a more precise inference of the detection range, it would be necessary to consider well-modeled sources and their known spatial distribution over the Galaxy. 

\begin{figure}[ht]
    \includegraphics[width=\linewidth]{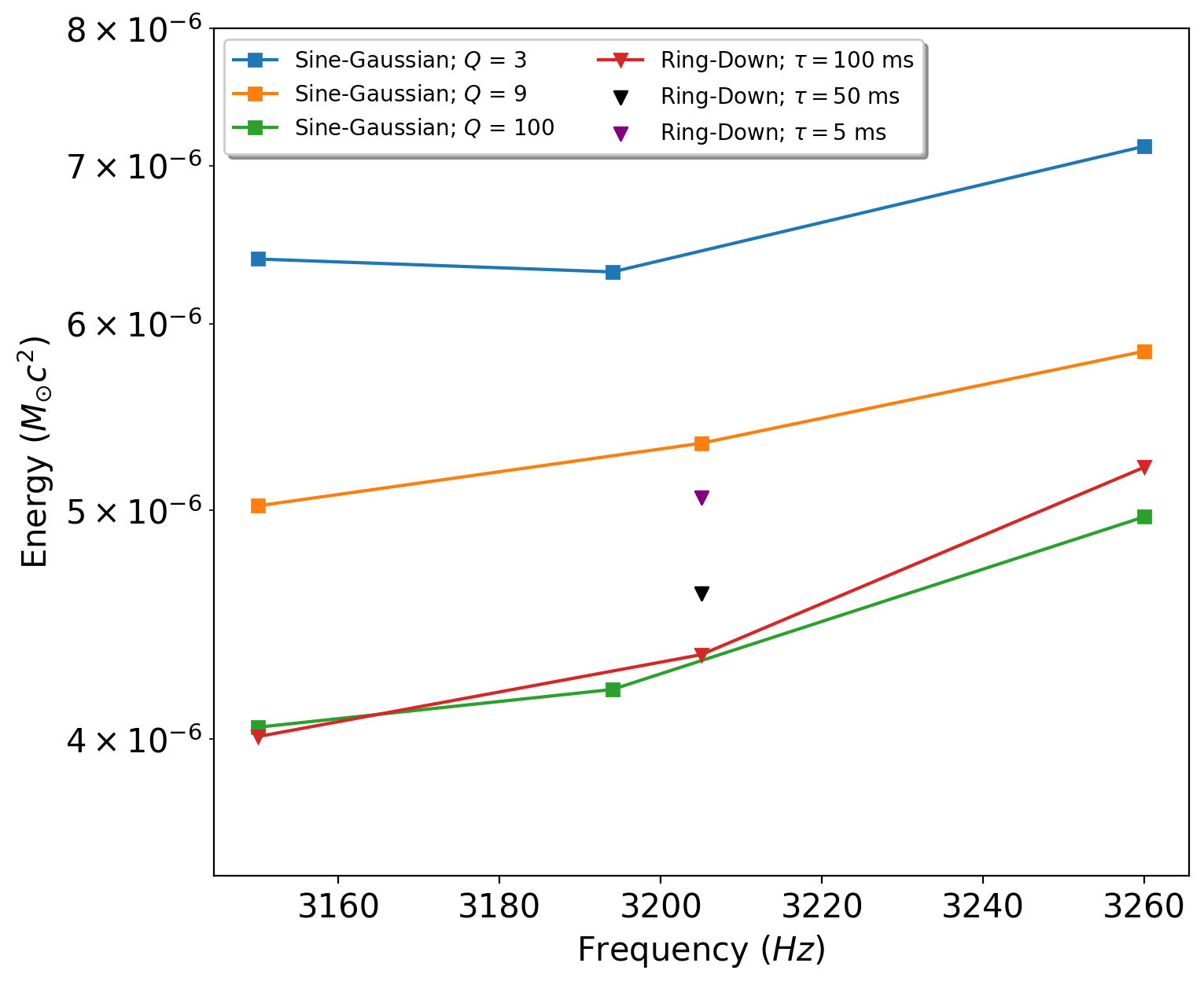}
    \caption{The emitted energy of the gravitational-wave burst expressed in units of solar masses which correspond to 50\% detection efficiency at an iFAR $\geq$ 100 years, for standard-candle sources emitting at 10 kpc for the waveforms listed in table~\ref{injections_hrss} into the Schenberg band. 
    }
    \label{energy_freq}
\end{figure}

Figure~\ref{energy_freq} illustrates the isotropically emitted energy from the gravitational-wave bursts as a function of the central frequency of the injections in the O3 data. The values shown correspond to the average $h_\text{rss}^{50\%}$ among the O3a and O3b epochs, with a 50\% detection efficiency and an iFAR $\geq$ 100 years. The progressive increase in energy observed in the figure is a consequence of the $f_{0}^2$ term in Equation~\ref{energy_iso} and also by the growth of the noise spectral amplitude over the frequencies. 

\begin{figure}[ht]
    \includegraphics[width=\linewidth]{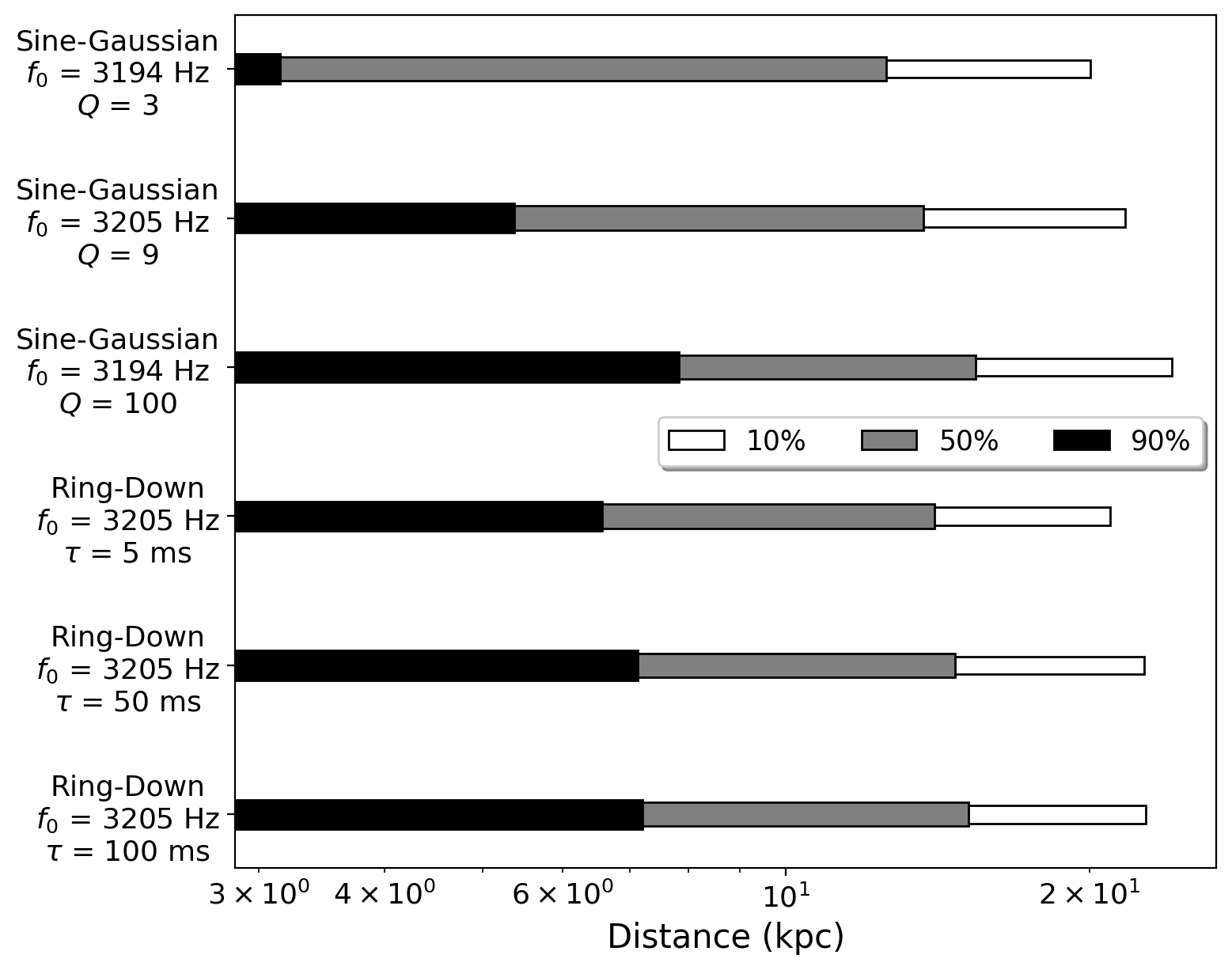}
    \caption{The minimum distance, in kiloparsecs (kpc), for the waveforms listed in Table~\ref{injections_hrss} with $f_0$ centered on the Schenberg band, for detection efficiencies of 10\%, 50\%, and 90\% at an iFAR $\geq$ 100 years and $E^{iso}_{GW} = 1\times10^{-5}M_{\odot}c^2$.}
\label{range_injections}
\end{figure}

Figure~\ref{range_injections} displays a range of distances for 10\%, 50\%, and 90\% detection efficiency ($d^{10\%}$, $d^{50\%}$, and $d^{90\%}$), represented by overlapping bars. These distances are calculated by Equation~\ref{energy_iso}, for a fixed energy of $E^{iso}_{GW} = 10^{-5}M_{\odot}c^2$, and they represent the range for the detection of gravitational-wave burst sources with isotropic emission. The range of detection distances scales with $\sqrt{E^{iso}_{GW}}$.

\subsection{Upper limits}
\label{subsec:upper_limits}

In the absence of an event with sufficient significance to indicate a detection, the efficiency results suggest other astrophysical interpretations. For example, GW bursts with low-strain amplitudes must occur more frequently to be detected, while bursts with high-strain amplitudes are more likely to be detected even if they occur less frequently. Based on the search sensitivity, it is feasible to set upper limits on the rate of GW burst using the Poisson distribution of potential astrophysical sources and link them to the strain amplitude. 

\begin{figure}[ht]
    \includegraphics[width=\linewidth]{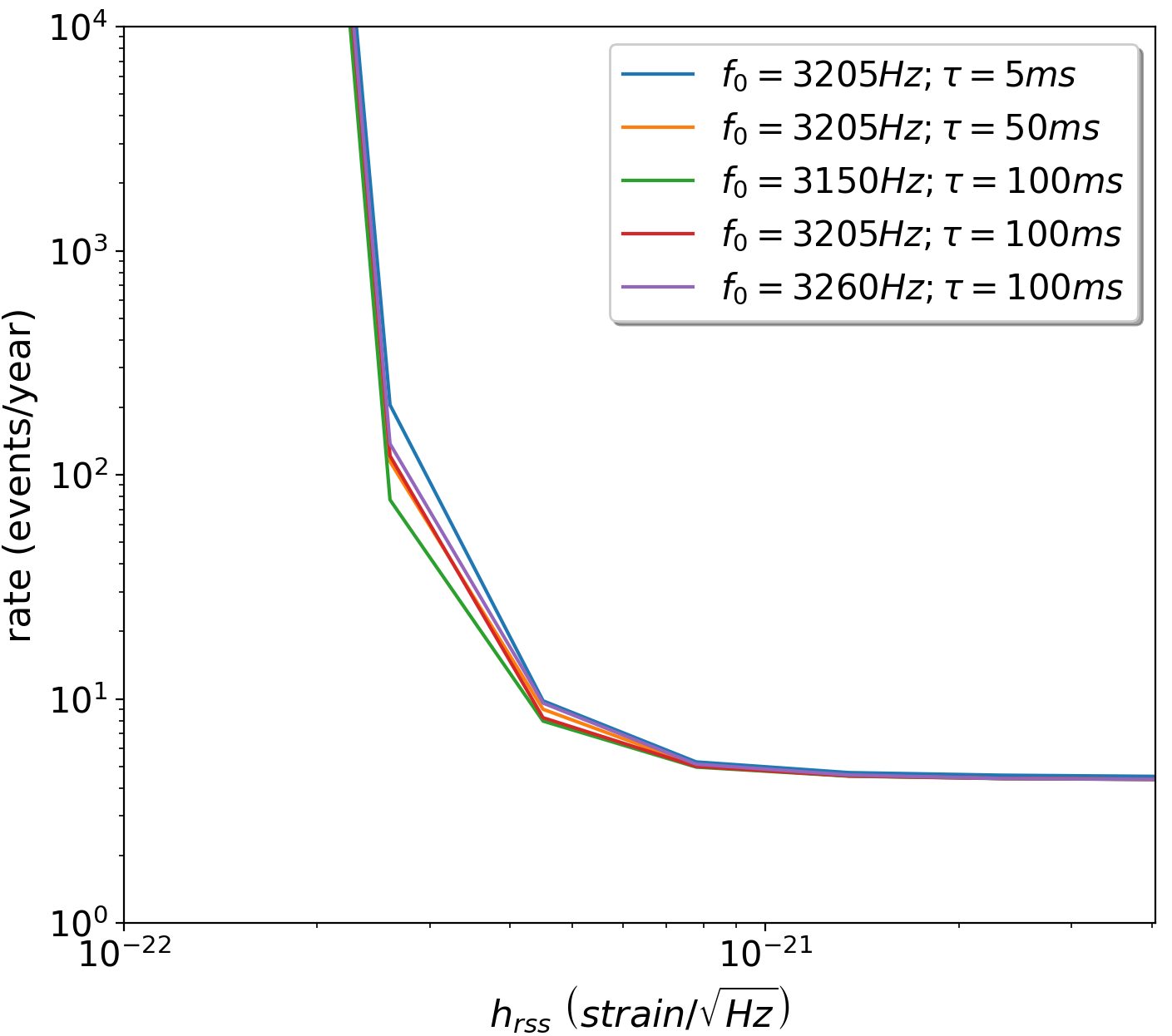}
    \caption{Upper limits of gravitational-wave burst event rate at 90\% confidence as a function of $h_\text{rss}$ for Ring-Down waveforms with central frequency in the Schenberg band [3150-3260] Hz. The results include both O3a and O3b epochs.}\label{upper_limit_RD}
\end{figure}

\begin{figure}[ht]
    \includegraphics[width=\linewidth]{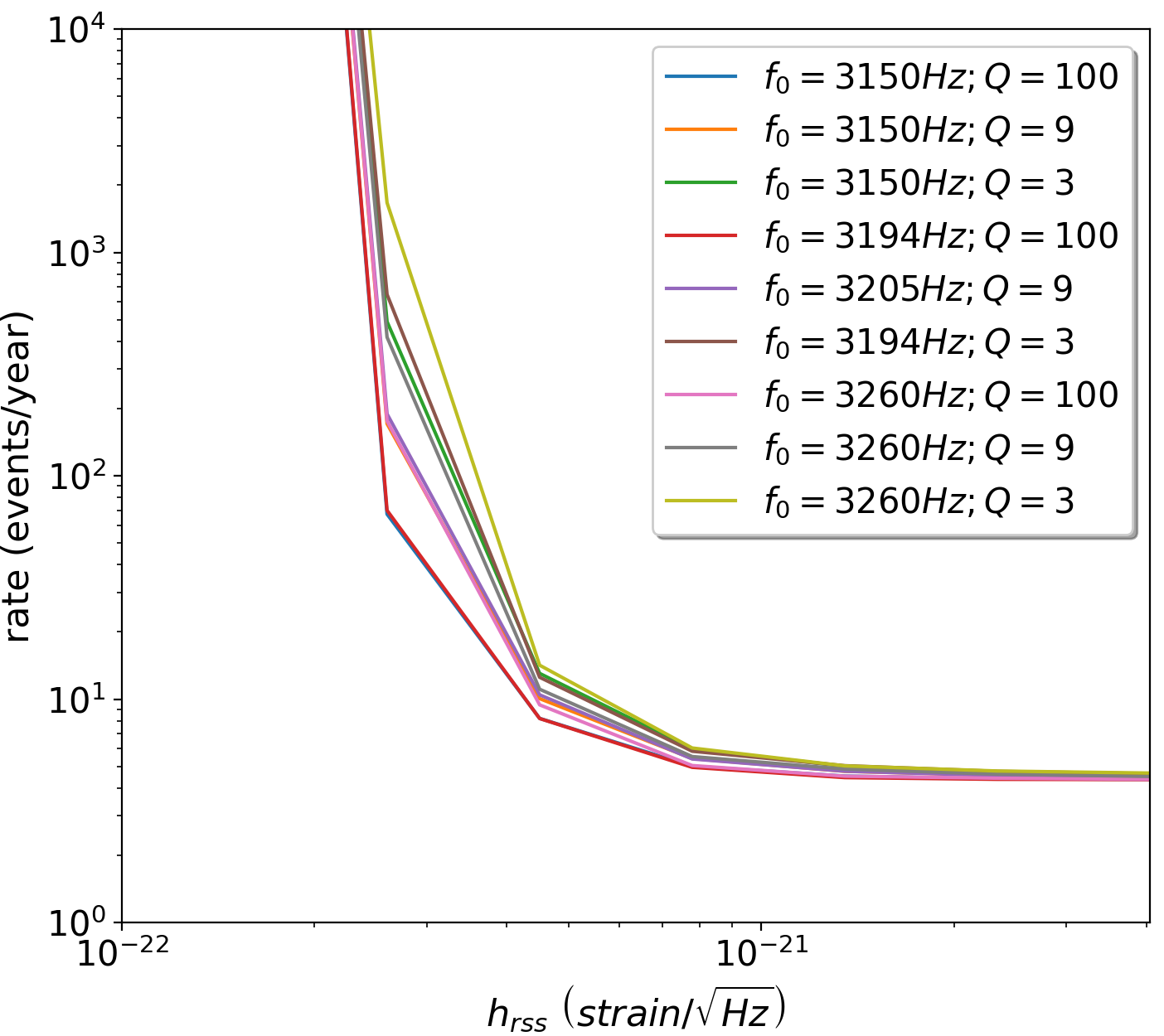}
    \caption{Upper limits of gravitational-wave burst event rate at 90\% confidence as a function of the strain amplitude $h_\text{rss}$ for Sine-Gaussian waveforms with central frequency in the Schenberg band [3150-3260] Hz.}\label{upper_limit_SG}
\end{figure}

Under the assumption of a Poisson distribution of random independent events (injections) in the livetime $T$, the upper limit on the total event rate at 90\% confidence level (C.L.) is given by: 

\begin{equation}
R_{90\%} = \frac{2.3}{\epsilon T},
\label{Rate_upper_limit}
\end{equation} 

\noindent where $2.3 =  -ln(1 - 0.9)$. In this case, the denominator is $\sum_i \epsilon_i T_i$ where the index $i$ indicates that the values of detection efficiencies $\epsilon$ and zero-lag livetime $T$ corresponds to the O3a and O3b trials. Further mathematical details concerning event rate upper limits can be found in \cite{upper_limit,upper_limit3, upper_limit2}. In the limit of strong signals ($\epsilon_{i} \approx 1 $) the quantity $\sum_i \epsilon_i T_i$ goes to 195.9 days, resulting in a 90\% confidence upper limit rate of $4.29$ $yr^{-1}$ in the [3150-3260] Hz band. Figures~\ref{upper_limit_RD} and~\ref{upper_limit_SG} show these upper limits as a function of signal strength for Ring-Down and Sine-Gaussian waveforms, respectively. These upper limits can be used to constrain the rate of gravitational-wave events originating from a known source population and discern their specific amplitudes.

\subsection{Revisiting the detectability of f-modes}
\label{subsec:f-modes}

Non-radial oscillations within neutron stars (NSs) can produce gravitational-wave bursts, with fundamental modes (f-modes) playing a critical role in this phenomenon \cite{f-modes_1, f-modes_revisited}. Depending on the equation of state (EoS) governing the neutron star matter, these f-modes can emit gravitational waves within the Schenberg band \cite{f_modes_EOS}. To assess the detectability of the f-modes, we adopted a similar methodology as in~\cite{Schenberg_fmodes}, considering the current search sensitivity characterization and the estimated detection range for aSchenberg. Nonetheless, due to uncertainties surrounding the oscillation mode mechanism and the highly dense matter physics in neutron stars, certain approximations become necessary.

\begin{figure}[ht]
\centering
\includegraphics[width=\linewidth]{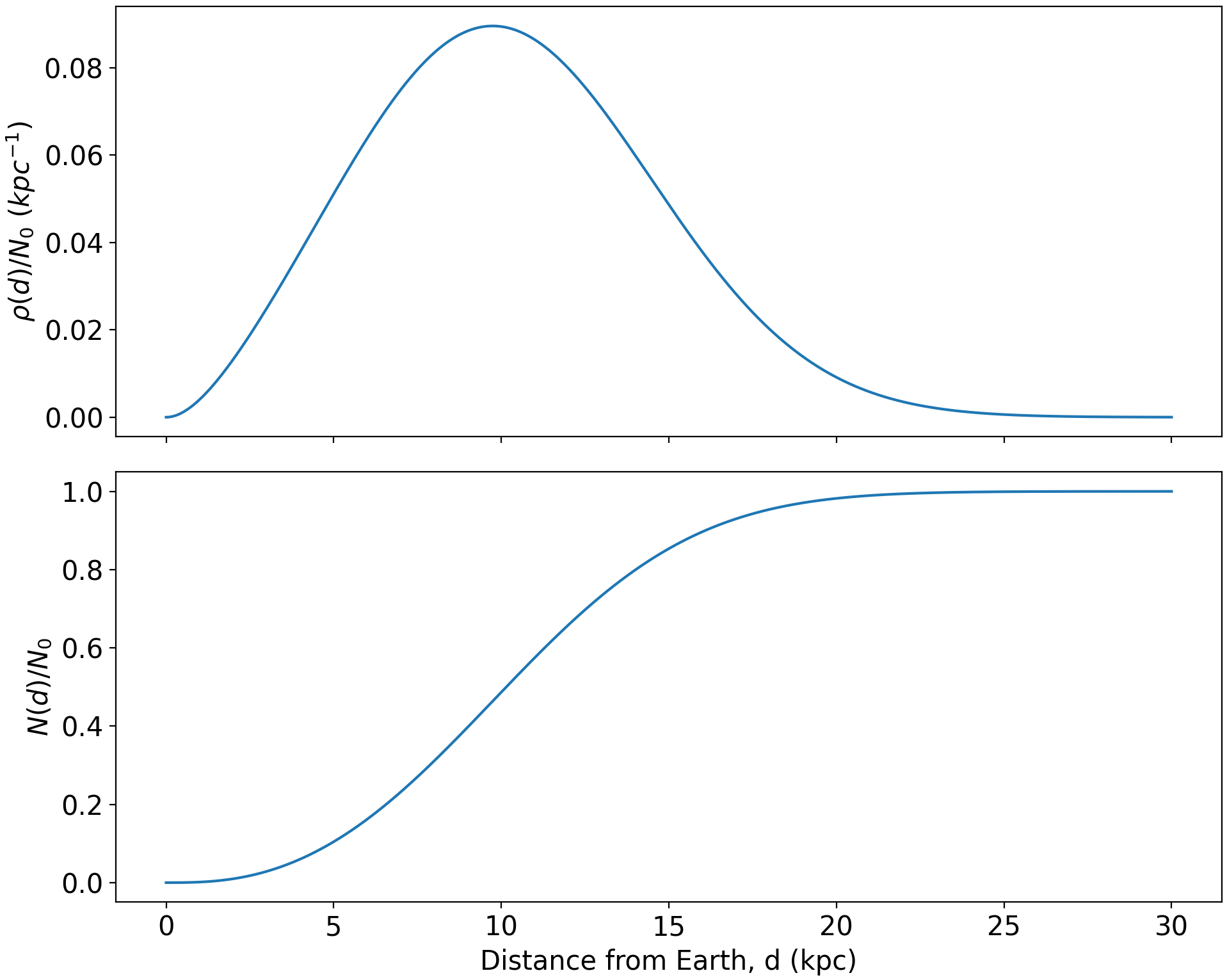}
\caption{The top subplot shows the normalized spatial distribution $\rho(d)/N_0$ plotted against the distance $d$. To estimate the total number of neutron stars at a specific distance $d$, the spatial distribution $\rho(y)$ can be integrated from 0 to $d$, resulting in the cumulative distribution function $N(d)$ shown in the bottom subplot.}
\label{fig:galactic_ns}
\end{figure}

We assume that f-modes from neutron stars are generated by the same mechanism as pulsar glitches and are excited to an energy level $E_{GW}$ supplied from the total energy of the glitch $E_{glitch}$. Typically, these gravitational wave signals exhibit a damping time of approximately 100 ms~\cite{f_modes_EOS}. Hence, the expected f-mode waveform morphology can be associated with Ring-Down injections, particularly with central frequencies of $f_0 = 3205$ Hz and a decay time ($\tau$) of 100 ms \cite{ringdown_eq_fmode}. To determine the detection range, Equation \ref{energy_iso} was solved for $d^{50\%}$ while considering various energy values. We focus on the Galactic detection range to estimate the detection potential of f-modes using a spatial distribution of neutron stars. The model adopted is based on the star formation pattern in the Galactic disk~\cite{Galatic_NS_model}, synthesized by the function: 

\begin{eqnarray}
\rho(d) = \frac{N_0 d^2}{\sigma^2_r z_0} \int_0^1  \exp\left[-\frac{xd}{z_0}\right]I_0 \left[\frac{R_e d \sqrt{1-x^2}}{\sigma_r^2} \right]\nonumber\\
\times  \exp\left[-\frac{R_e^2 + d^2(1-x^2)}{2\sigma^2_r} \right] dx,
\label{NS_dist_model}
\end{eqnarray}

\noindent where $I_0$ is the modified Bessel function, $\sigma_r = 5$ kpc is a radius parameter, $N_0$ is the total number of Galactic neutron stars, $z_0 = 2.0$ kpc is the adopted disk thickness, $R_e = 8.25$ kpc is the distance from the Galactic Center to Earth, and the scaled variable $x$ is related to the height $z$ (cylindrical coordinates) by $x = z/d$. Further details regarding this Galactic neutron star distribution model can be found in~\cite{Galatic_NS_model}. Figure~\ref{fig:galactic_ns} presents both the normalized spatial distribution and the cumulative distribution function of Galactic neutron stars at a distance of 30 kpc from Earth.

The glitch rate can be estimated by observing the known population of neutron stars, the pulsars~\cite{NS_glitches_review}. However, not all studied pulsars exhibit glitches. The Jodrell Bank Observatory (JBO) timing program monitors 800 pulsars, and although in some cases it has collected over 50 years of timing history on individual objects, only 178 pulsars have at least one detected glitch~\cite{glitch_rate}. It is likely that glitches will occur in pulsars without any currently known events, suggesting that the fraction of glitching pulsars (approximately 20\%) should be considered a lower limit of the intrinsic fraction. 

The pulsars have different features on glitch rates and indicate a dependence on their spin-down rate and characteristic age of the pulsar~\cite{glitch_age_Lyne, glitch_age_Espinoza}. The physical parameters dependence and the limitations on monitoring time make it challenging to characterize the average rate of glitches for the entire population of pulsars. In~\cite{glitch_rate} the rate ($R_g$) for each glitching pulsar is calculated as constant in time and, therefore, should be seen as only an approximation. 
To avoid some overestimation of the glitch rate, the authors considered the average glitch rate for the entire interval in which the pulsar has been monitored and not the interval between the first and last detected glitches. The $R_g$ $>$0.05 yr$^{-1}$ was adopted as a lower bound since the extrapolation to lower rates is greatly hindered by the total observing time for these 134 samples.

According to \cite{pulsar_population}, the estimated population of pulsars in the entire Galaxy, considering the correction for the beaming factor, is approximately $2.4 \times 10^5$ pulsars with a luminosity greater than 0.1 mJy kpc$^2$ at 1400 MHz. By integrating Equation~\ref{NS_dist_model} within the detection distance $d = d^{50\%}$ and $N_0 \sim 2.4 \times 10^5$, we can determine the minimal number of glitching pulsars that could be potentially observed by the aSchenberg and LIGO detectors. If we assume that all Galactic neutron stars exhibit the same glitching behavior as the known pulsars in terms of mean glitch rate and the fraction of the population that experiences glitches, then $N_0$ would be $10^8$ \cite{Galatic_NS_model} increasing the event rate for the same emitted energy. However, it is important to note that this is a speculative assumption as the non-pulsar population includes many older neutron stars, and glitch occurrence reduces as pulsars age and for those with low spin-down rates \cite{glitch_rate}.

\begin{figure}[ht]
\includegraphics[width=\linewidth]{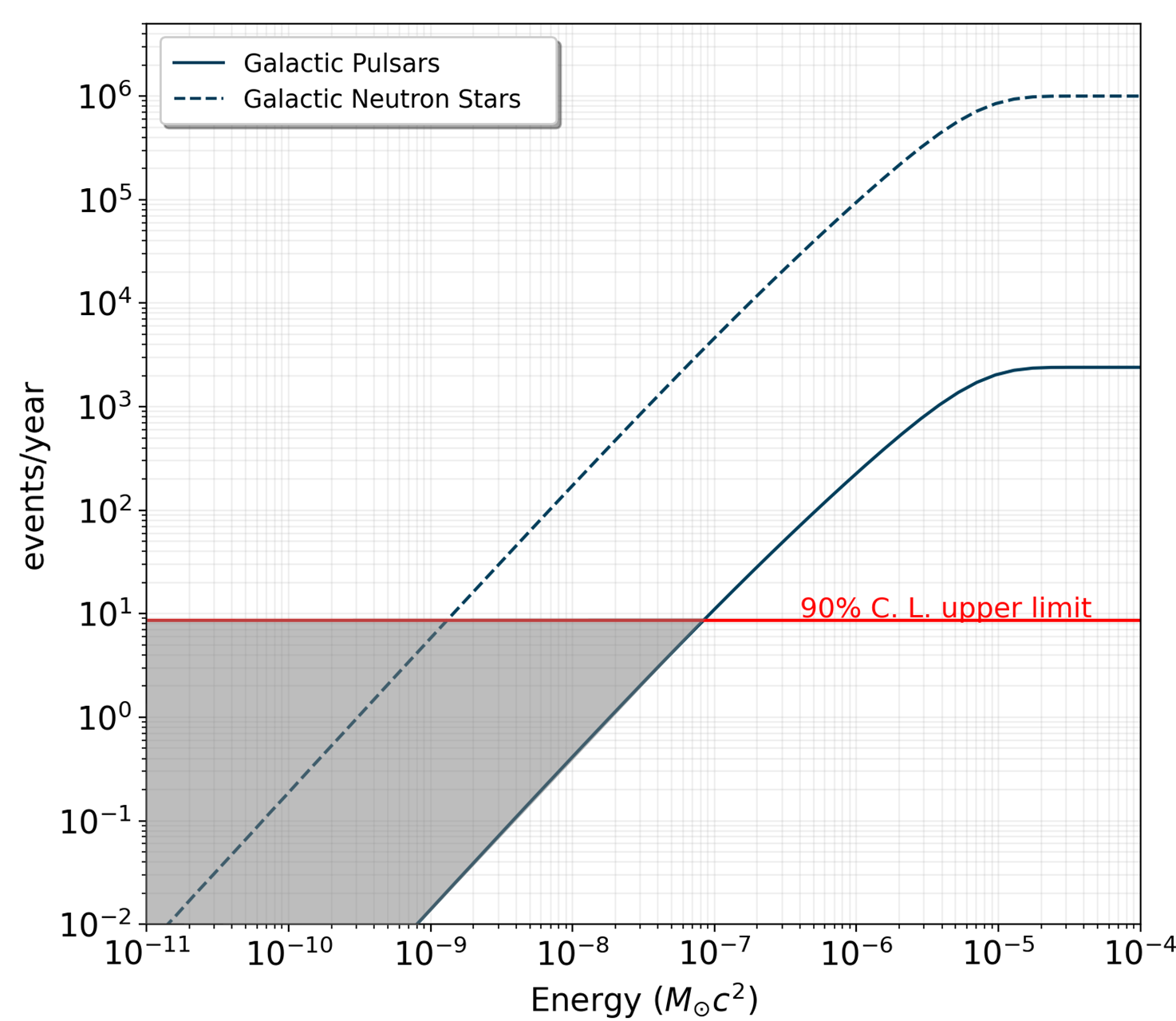}
\caption{
Number of f-mode events per year as a function of glitch energy for the estimated population of Galactic pulsars and neutron stars. The Galactic Neutron Stars curve shows the hypothetical scenario of equal glitch activity from the pulsar population. The NS/pulsar rate is estimated using the lower limit of individual glitch rate ($R_g = 0.05 $  yr$^{-1}$) and assumes $20\%$ of the pulsar population presenting glitches. As previously clarified, these restrictions are imposed by the monitoring time of the pulsars, resulting in a conservative rate of f-modes. On the other hand, the rate of f-modes is constrained by the upper limit value (with $90\%$ C.L) for the burst rate at $50\%$ detection efficiency, which is 8.6 events per year (red line). The grey region bounded by these curves shows the possible range of f-mode rates for different glitch energies, up to a maximum energy of $\sim 8 \times 10^{-8} M_{\odot}c^2$ for the pulsars population.}\label{rate_vs_energy}
\end{figure}

Figure~\ref{rate_vs_energy} illustrates the potential number of f-mode events per year for different values of $E_{glitch} \approx E_{GW}^{iso}$, considering the estimated population of Galactic pulsars and neutron stars at a distance of $d^{50\%}$. If we assume that the entire population of glitches in the Galaxy exhibits an energy $E_{glitch} \approx E_{GW}^{iso}$, this figure can be interpreted as a range of values for the f-mode detection rate in the aSchenberg and LIGO at 3205 Hz. 

The energy range of gravitational waves emitted by NS glitches, $E_{GW} \approx E_{glitch} \approx 10^{37} - 10^{44}$ erg $\approx 10^{-17} - 10^{-10} M_{\odot}c^2$, is quite broad, reflecting the wide range of glitch sizes \cite{GW_fmode2021}. By analyzing Figure~\ref{rate_vs_energy} for this energy range, it becomes evident that f-modes excited by NS glitches with those energies would be detected with a low rate, making them a not very promising source of GW bursts for the aSchenberg. Such energy range gives rates significantly lower than the upper limit for the gravitational-wave burst event rate at a 90\% confidence level, as expected. For Ring-Down injections at $f_0 = 3205$ Hz and $\tau = 100$ ms, the upper limit is $\sim$8.6 events per year at a 50\% detection efficiency, as shown in Figure~\ref{upper_limit_RD}. 

The results obtained here are considerably less optimistic for the detectability of f-modes by the aSchenberg spherical antenna compared to~\cite{Schenberg_fmodes}, which assumed a detection requirement of signals with SNR = 1 and SNR = 3. Despite considering an intrinsic glitch rate approximately five times higher than~\cite{Schenberg_fmodes} and accounting for a better strain sensitivity of the Schenberg antenna (the ultimate sensitivity), our criteria for a detection range only for significant signals (with iFAR $\geq 100$ years) significantly reduces the expected detection rate.

\section{SUMMARY AND DISCUSSION}
\label{sec:conclusion}

This work presents an all-sky search for transient gravitational waves of short duration in LIGO data from the third Advanced LIGO/Virgo run (O3) without assuming specific signal morphologies, with the aim of characterizing the detectability of burst signals by the aSchenberg antenna, assuming its similarity in sensitivity to LIGO during O3. Despite the current inactivity of the Brazilian antenna, we investigated its detection potential based on the available O3 data. The LIGO Livingston and LIGO Hanford operated with improved sensitivity from April 2019 to March 2020. Offline analysis of the collected O3 data was performed using the coherent WaveBurst pipeline covering frequencies from 512 Hz to 4069 Hz. Gravitational-wave burst candidates were selected based on their reconstructed bandwidth overlapping with the Schenberg band.

A total of 195.9 days of LIGO coincidence observation time met the data quality requirements for our analysis. To account for differences in the background event count distribution, the data were divided into three separate chunks. However, no significant burst event candidates were found in any of the chunks. The detected signals were determined to be accidental coincidences according to the expected significance distribution for background events. Nevertheless, this result does not imply the Schenberg antenna would not detect any signals if operated with the same sensitivity at that frequency. Given the random nature of events in the universe, observing zero events is consistent with a nonzero event rate under Poisson statistics.

Regarding the search sensitivity to circularly polarized waveforms, we evaluated the efficiency of the cWB algorithm in recovering expected gravitational-wave burst waveforms from all-sky directions using an iFAR threshold of 100 years. We focused on two simple waveform morphologies, namely Sine-Gaussians with varying quality factors ($Q = 3, 9$ and $100$) throughout the Schenberg band and Ring-Downs with specific damping times ($\tau = 5, 50$ and $100$ ms). By studying the search sensitivity for simulated signals, we characterized the detection range of the aSchenberg antenna. We estimated the amount of mass that, when converted into gravitational-wave burst energy at a distance of 10 kpc, would be detectable with 50\% efficiency, assuming an isotropic emission pattern. This estimation yields an value of $E^{iso}_{GW} \approx 5 \times 10^{-6} M_{\odot}c^2$.

To provide further insights into the detection capability, we also estimated the distances at which representative waveforms could be detected with 10\%, 50\%, and 90\% efficiency, based on the gravitational-wave energy. The detection range analysis employed an isotropic emission pattern to generalize the results. A more precise detection distance for a well-modeled event can be determined by considering the non-isotropic nature of gravitational-wave energy emission as well as the potential increase in detection range for an optimally oriented source concerning the line of sight.

We also established upper limits at a 90\% confidence level on the rate of gravitational-wave bursts reaching Earth. By studying the efficiency of the detection process, we associated these upper limits with the strain amplitude ($h_\text{rss}$) of the selected waveforms. This approach enables a more accurate constraint on the population study of burst sources within the Schenberg band.

Furthermore, a Ring-Down injection with $\tau = 100$ ms was used to investigate f-mode unstable neutron stars, triggered by glitches, as sources of bursts for the Brazilian antenna. Using a simple Galactic neutron star distribution model, along with lower limits on the fraction of pulsars that exhibit glitches and intrinsic glitch rates, we conducted a preliminary assessment of the detectability of f-modes presented in Figure \ref{rate_vs_energy}. The region constrained by this curve and the upper limits on the rate of bursts gives the possible range of f-mode rates for different glitch energies, up to a maximum glitch energy of approximately $8 \times 10^{-8} M_{\odot}c^2$ for the pulsar population.

Our analysis focused on f-modes, which are expected to be the most common source of short gravitational-wave transients in the Schenberg band. Although the fundamental mode contains the main part of the energy among the quasinormal modes, their energy levels when triggered by neutron star glitches are generally low, suggesting that they are not a significant source of GWs for aSchenberg. However, the excitation of f-modes in isolated neutron stars can be triggered by other physical processes that result in the emission of GW transients with higher energies \cite{fmode_perspective}. These mechanisms are typically related to stellar oscillations, such as starquakes. 

The magnetar giant flares may be associated with these starquakes \cite{magnetar_review}. During a magnetar giant flare, the magnetar experiences a catastrophic rearrangement of its interior magnetic field, leading to the emission of GWs. The highest estimated energies of GWs emitted during magnetar giant flares can reach the order of $5 \times 10^{-6} M_{\odot} c^2$ \cite{magnetar_maxEnergy}, corresponding to a Galactic detection range within the sensitivity of aSchenberg ($\sim$10 kpc), assuming the signal central frequency is within the aSchenberg band. The rate of giant flares per magnetar has been estimated to be $\lesssim0.02$ yr$^{-1}$ \cite{magnetar_maxEnergy}. Considering the presence of at least 20 known Galactic magnetars within the detectable range \cite{magnetar_catalog}, we can anticipate $\sim$0.4 potential events per year. Despite the promising event rate, GW signals from magnetar giant flares have not yet been detected in the data from Advanced LIGO and Virgo \cite{magnetar_search_O2, magnetar_search_O3}.

Gravitational waves from core-collapse supernovae (CCSN) cover frequencies ranging from several hundred Hz to a few kHz and are also a feasible source of GW bursts in the Schenberg band \cite{CCSNe_waveform}. Its estimates of the energy radiated as GWs tend to be conservative, suggesting a total release not exceeding $10^{-6} M_{\odot} c^2$ \cite{CCSNe_energy}, with only a small fraction occurring in the Schenberg band. The estimated rate of Galactic supernovae is 4.6$^{7.4}_{-2.7}$ events per century \cite{Galactic_SN_rate} and the current GW energy constraint (based on O3 data) is $10^{-4} M_{\odot} c^2$ at 50 Hz \cite{constrain_CSSN_energy}.

The f-mode also plays an important role in proto-neutron star (PNS) oscillations, formed during the post-collapse evolution of core-collapse supernovae and carrying away most of the GW power during this process \cite{energy_f-mode_CSSN}. However, when hot PNS produces f-modes they generate GWs with lower frequencies than those of their cold, old descendants \cite{f-modes_CCSN}, and their signal does not achieve the Schenberg band. Given the dependence on the average density of the star, the f-mode frequencies increase as the proto-neutron star core shrinks driven by neutrino losses. In such sources, the components of the signal within the Schenberg band are characterized by a "haze" emission correlated with the phase of violent accretion flows onto the PNS, although the origin of this haze is still a bit unclear \cite{energy_f-mode_CSSN}.

The eventual observation of f-modes would give valuable hints about the physical conditions of extremely compact matter in neutron stars and the EoS would be strongly constrained~\cite{EoS_constrain}. Aligned to our work, the authors from \cite{f_mode_Dixeena} give a detailed discussion of upper limits in O3 data for f-modes triggered by glitches in NSs with frequencies between $2.2 - 2.8$ kHz, based on the adopted EoS. The study focuses on the prospects for detecting f-mode emission from different conditions of intrinsic parameters (frequency and damping time related to mass and EoS) and extrinsic parameters (sky direction and orientation of the source) of the neutron stars. 

Our study aims to provide a scientific framework for characterizing the feasibility of reconstructing the Schenberg antenna, taking into account the state-of-the-art approaches to burst detection. However, it is important to note that our work does not provide a definitive assessment of the future of the project. Achieving the \textit{ultimate} sensitivity of the Schenberg detector would require significant research and development efforts in various technological and experimental aspects, which are beyond the scope of our study.

To comprehensively assess the detection potential of the Schenberg antenna, it is essential to explore various categories of gravitational wave signals present in the O3 data. One important aspect is the investigation of continuous signals \cite{continuouswave} and signals originating from compact binary coalescence. In the context of continuous waves, relativistic calculations incorporating realistic hadronic equations of state have shown the maximum spin rates of neutron stars can result in the emission of gravitational wave signals above 2 kHz \cite{max_spin}. Furthermore, the existence of strange stars, characterized by the formation of quark matter at extreme densities \cite{strange_star,NS_review, evidence_Strange_matter}, represents another intriguing source of gravitational waves within the Schenberg frequency band. These exotic objects can exhibit higher rotational frequencies compared to neutron stars \cite{strange_star_rot}. 

Additionally, previous studies have explored the detectability of primordial black hole binaries known as massive astrophysical compact halo objects (MACHOs) using the Schenberg antenna \cite{MACHO}. These investigations focused on assessing the sensitivity of the antenna to black hole binaries with individual masses of $0.5 M_{\odot}$ just before the merger. During the final inspiral phase, these systems generate short-duration transients within the frequency range of [3150-3260] Hz. 

In addition to the search for gravitational-wave bursts at the Schenberg sensitivity range, a thorough investigation of the potential improvements in data quality through the inclusion of a resonant antenna in the network of detectors is crucial. The Schenberg's impact in this field is not solely based on its own detection abilities, but rather on its position within a network of detectors. This examination can shed light on the enhanced capabilities and scientific benefits such an addition may bring to gravitational wave astronomy.

\begin{acknowledgments}
The authors would like to thank Marco Drago, Shubhanshu Tiwari, and Dixeena Lopez for stimulating discussions and support with cWB software. 
ODA would like to acknowledge \textit{Funda\c{c}\~ao de Amparo \`a Pesquisa do Estado de S\~ao Paulo} (FAPESP) for financial support under the grant numbers 1998/13468-9, 2006/56041-3, the Brazilian Ministry of Science, Technology and Inovations and the Brazilian Space Agency as well.
Support from the \textit{Conselho Nacional de Desenvolvimento Cientifíco e Tecnológico} (CNPq) is also acknowledged under the grants number 302841/2017-2, 310087/2021-0. This study was financed in part by the \textit{Coordenação de Aperfeiçoamento de Pessoal de Nível Superior} - Brasil (CAPES) - Finance Code 001. This material is based upon work supported by NSF’s LIGO Laboratory which is a major facility fully funded by the National Science Foundation. The authors are grateful for computational resources provided by the LIGO Laboratory	(CIT) and supported by National Science Foundation Grants PHY-0757058 and PHY-0823459.

\end{acknowledgments}

\appendix

\nocite{*}

\bibliography{references}

\end{document}